\documentclass[onecolumn,draft]{article}

\newlength{\defbaselineskip}
\usepackage{psfig,pstricks,pst-grad,color,fancybox,graphics} %multicol
\usepackage{latexsym}
\usepackage{amsmath}
\usepackage{amssymb}
\usepackage{enumerate}
\usepackage{bbm}

\begin{document}

\title{\bf Parametric Oscillation with Squeezed Vacuum Reservoirs }

\author{ A. Mebrahtu$^{1,2}$\\ 
$^1$ Institut f{\"u}r Theoretische Physik, Universt{\"a}t Hannover,\\ 
Appelstra{\ss}e 2, D-30167 Hannover, Germany\\
$^2$Department of Physics, Mekelle University, P.O.Box 231, Mekelle, Ethiopia}

\maketitle

{\noindent {\bf Abstract:} Employing the quantum Hamiltonian describing
the interaction of  two-mode light (signal-idler modes) generated by a 
nondegenerate parametric oscillator (NDPO) with two uncorrelated squeezed
vacuum reservoirs (USVR), we derive the master equation. 
The corresponding Fokker-Planck equation for the
Q-function is then solved employing a propagator method developed by 
K. Fesseha [J. Math. Phys. {\bf 33} 2179(1992)]. 
Making use of this Q-function, we calculate the quadrature 
fluctuations of the optical system. From these results we infer that the 
signal-idler modes are in  
squeezed states. 
When the NDPO operates below threshold we show that, for a large 
squeezing parameter, a squeezing amounting to a noise suppression 
approaching 100\% below the vacuum level in the of the quadratures can be 
achieved.\\

\section {Introduction}

Nonclassical effects of light such as squeezing, antibunching and 
sub-Poissonian statistics have been attracting the attention of several 
authors in quantum optics over the last 
decades ~\cite{1,2, 3, 4, 5, 6, 7, 8, 9}. A review article on nonclassical  
states of the first 75 years is found in Ref.~\cite{2}. 

Squeezed states are nonclassical states characterised by a 
reduction of quantum fluctuations (noise) in one quadrature component  below 
the vacuum level (quantum standard limit), or below that achievable in a 
coherent state ~\cite{4, 6} at the expense of 
increased fluctuations in the other component such that the product of these 
fluctuations still obeys the uncertainty relation ~\cite{4,5}. 

It was Takahashi ~\cite{10} who, in 1965, first pointed out 
that a degenerate parametric amplifier enhances the noise in one 
quadrature component  and attenuates it in the other quadrature. 
This prediction has been confirmed by several authors 
for degenerate and nondegenerate parametric amplifiers and oscillators. 
Operating below threshold, the parametric amplifier is a 
source of squeezed states. In the initial 
experiments carried out to observe squeezing, a noise reduction of 4-17\% 
relative to the quantum standard limit has been obtained ~\cite{11}. In order 
to increase the gain, the parametric medium may be placed inside an optical 
cavity where it is coherently pumped and becomes a parametric 
oscillator ~\cite{12, 13, 14, 15, 16, 17}. 

An optical parametric oscillator is a quantum device with a definite threshold 
for self sustained oscillations. It is one of the most interesting and well 
characterised optical devices in quantum optics. 
This simple dissipative quantum system plays an important role in the study of 
squeezed states. In a parametric oscillator a strong pump photon interacts 
with a nonlinear-medium (crystal) inside a cavity and is down-converted into 
two photons of smaller frequencies. 
In the NDPO we assume that the strong pump photon is down converted into two 
modes and these modes are referred to as signal and idler modes.

A quantum-mechanical treatment of different optical systems such as the NDPO 
is essential as they may generate squeezed states with nonclassical properties 
which have potential applications in quantum optical 
communications ~\cite{4} and computation~\cite{19}, gravitational wave 
detection ~\cite{19, 20, 21, 22}, interferometry~\cite{22, 23, 24}, 
spectroscopical measurements ~\cite{25} and for the study of fundamental 
concepts.

For systems with nonclassical features such as the NDPO, for which the 
Glauber-Sudarshan P-function is highly singular ~\cite{8, 26, 27}, one 
may use the Q-function. The Q-function is expressible in terms of the 
Q-function propagator and the initial Q-function. It is possible to determine 
the Q-function propagator by directly solving the Fokker-Planck equation. 
In this paper, we find it convenient to  evaluate the Q-function
propagator applying the  method developed in ~\cite{28}.

The main aim of this paper is to calculate 
the amount of squeezing that could be generated by the NDPO coupled to two 
USVR with the help of the Q-function. We show that squeezing 
of the output may be optimised and reach 100\%.

\section{ The Master Equation}
 
The description of system-reservoir interactions via the master 
equation is a standard technique in quantum optics ~\cite{6, 7}. In this 
section, however, we found it useful to include a non-detailed a derivation 
of the master equation describing the interaction of the signal-idler modes 
generated by a NDPO coupled to two USVR in order to make the paper more 
self-contained. 

Denoting the density operator of the optical system and the squeezed reservoir 
modes by 
$ \hat{\chi}(t) $, the density operator for the system alone is defined by 
\begin{equation*}
 \hat{\rho}(t) ={\rm Tr}_R(\hat{\chi}(t)), 
 \end{equation*}
where $ {\rm Tr}_R $ indicates that the trace is taken over the reservoir 
variables only. The density operator $ \hat{\chi}(t) $ 
evolves in time according to 
\begin{equation}
 \frac{d \hat{\chi}(t)}{dt} = \frac{1}{i \hbar} \Big[ \hat{H}_{SR}(t), 
\hat{\chi}(t) \Big],  \label{1}
 \end{equation} 
where $\hat{H}_{SR}(t)$ is the Hamiltonian describing the interaction between 
the system and the reservoirs. 
Note that the Hamiltonian describing only the reservoir modes 
($\hat{H}_{R}(t)$) is not involved 
in the derivation as it cancels out when we apply the cyclic property of the 
trace. Furthermore, in oder to simplify our calculations, the Hamiltonian that 
describes the interaction of the system with the pump mode
($\hat{H}_{S}(t)$) will be added at the end of the derivation.

Since initially the system and the reservoirs are 
uncorrelated, one can write, for the density operator of the system and the 
reservoirs at the initial time ($t=0$), that 
$ \hat{\chi}(0) = \hat{\rho}(0)\otimes \hat{R}$ ~\cite{6}, where 
$\hat{\rho}(0)$ and $\hat{R}$ are the density operators of the system and the 
reservoirs at the initial time, respectively. Then in view of this relation, 
Eq.~(\ref{1}) results in 
\begin{equation}
\frac{d \hat{\chi}(t)}{dt} =\frac{1}{i \hbar} \Big[ \hat{H}_{SR}(t), 
\hat{\rho}(0)\otimes \hat{R} \Big] - 
   \frac{1}{\hbar^2}\int_0^t dt^\prime    
\Big[ \hat{H}_{SR}(t), \big[ \hat{H}_{SR}(t^\prime), \hat{\chi}(t^\prime)  
\big]\Big]. \label{2} 
\end{equation} 
Applying the weak coupling approximation which implies that 
$\hat{\chi}(t^\prime )= 
\hat{\rho}(t^\prime)\otimes \hat{R}$ ~\cite{6}, 
it follows that
\begin{equation}
 \frac{d\hat{\rho}(t)}{dt}=\frac{1}{i\hbar} 
{\rm Tr}_R \bigg\{\Big[ \hat{H}_{SR}(t), 
\hat{\rho}(0)\otimes\hat{R} \Big]\bigg\}  - \frac{1}{\hbar^2}\int_0^t 
dt^\prime {\rm Tr}_R\bigg\{\Big[ \hat{H}_{SR}(t), \big[\hat{H}_{SR}(t^\prime), 
\hat{\rho}(t^\prime) \otimes\hat{R}\big] \Big]\bigg\}. \label{3}
\end{equation}

We consider the system to be a two-mode light with frequencies 
$ \omega_a $ and $ \omega_b $ in a cavity  coupled to two USVR. 
The interaction between the two-mode light and the squeezed vacuum reservoirs 
can be described, in the interaction picture, by the Hamiltonian 
\begin{equation}
 \hat{H}_{SR}(t) = i\hbar\Big[ \sum_j \lambda_j \big( \hat{a}^{\dag}
\hat{A}_j \, e^{i(\omega_a - \omega_j)t} 
-\hat{a} \hat{A}^{\dag}_j \, e^{-i(\omega_a - \omega_j)t}\big) + 
\sum_k \lambda_k \big( \hat{b}^{\dag}\hat{B}_k \, e^{i(\omega_b - 
\omega_k)t} -\hat{b}\hat{B}^{\dag}_k \, e^{-i(\omega_b - 
\omega_k)t} \big) \Big], \label{4} 
\end{equation}
in which $ \hat{a}\,( \hat{a}^{\dag} ) $ and $ \hat{b}\,(\hat{b}^{\dag}) $ are 
the annihilation (creation) operators for the intracavity modes and 
$\hat{A}_j\,(\hat{A}^{\dag}_j)$ 
and $\hat{B}_k\,(\hat{B}^{\dag}_k)$ are the annihilation (creation) operators 
for the reservoir modes with frequencies $ \omega_j $ and $\omega_k$, 
respectively. The coefficients $\lambda_j$ and $\lambda_k$ are coupling 
constants describing the interaction between the intracavity modes and the 
reservoir modes. Applying the cyclic property of trace and the relation 
${\rm Tr}_R \big(\hat{R}\otimes\hat{H}_{SR}(t)\big)=
\langle \hat{H}_{SR}(t)\rangle_R $, and taking into account that,
for squeezed vacuum reservoirs~\cite{6}, 
\begin{equation*}
\langle \hat{A}_j\rangle_R=\langle \hat{A}^{\dag}_j\rangle_R = 
\langle \hat{B}_k\rangle_R  = \langle \hat{B}^{\dag}_k\rangle_R = 0, 
\end{equation*} 
one can show 
\begin{equation*}
\frac{1}{i\hbar} {\rm Tr}_R \bigg\{\Big[ \hat{H}_{SR}(t) , 
\hat{\rho}(0)\otimes\hat{R} \Big]\bigg\} = 0,  
\end{equation*}  
and as a result, expression (4) reduces to 
\begin{equation}
\frac{d\hat{\rho}(t)}{dt}=-\frac{1}{\hbar^2}\int_0^t\!dt^\prime\,
{\rm Tr}_R\bigg\{ \Big[ \hat{H}_{SR}(t), \big[\hat{H}_{SR}(t^\prime),
\hat{\rho}(t^\prime)\otimes\hat{R}\big]\Big]\bigg\}. \label{eqno 5}  
\end{equation} 
Applying the Markov approximation, in which $ \hat{\rho}(t^\prime)$ 
is replaced by $ \hat{\rho}(t), $ and using the cyclic property of the trace,  
the above equation can be expressed as 
\begin{eqnarray}
\frac{d\hat{\rho}(t)}{dt}
 &=& -\frac{1}{\hbar^2}\int_0^t\!dt^\prime 
\Big[\langle \hat{H}_{SR}(t)\hat{H}_{SR}(t^\prime)\rangle_R 
\hat{\rho}(t) -\langle \hat{H}_{SR}(t^\prime)\hat{H}_{SR}(t)\rangle_R
\hat{\rho}(t) \nonumber\\
 & & -\hat{\rho}(t) \langle\hat{H}_{SR}(t) \hat{H}_{SR}(t^\prime)\rangle_R
 + \hat{\rho}(t) \langle\hat{H}_{SR}(t^\prime)\hat{H}_{SR}(t)\rangle_R \Big].
 \label{6}
\end{eqnarray}  
We note again that for squeezed vacuum reservoirs~\cite{6}
\begin{subequations}
\begin{align}
&\langle\hat{A}_j\hat{A}_l\rangle_R = -M_{A}\delta_{l,2j_a-j}, \\
&\langle\hat{A}^{\dag}_j\hat{A}_l\rangle_R = N_{A}\delta_{j,l}, \\
&\langle\hat{A}_j\hat{A}^{\dag}_l\rangle_R = (N_A+1)\delta_{j,l},  
\end{align}
\label{7}
\end{subequations}
where $\delta_{j,l}$ is the Kronecker delta symbol and
\begin{eqnarray}
\langle\hat{A}_j\hat{B}_m\rangle_R =   \langle\hat{A}_j\hat{B}^{\dag}_m
\rangle_R  = \langle\hat{A}^{\dag}_j\hat{B}^{\dag}_m\rangle_R=0. \label{8} 
\end{eqnarray}
This equation is a consequence of the fact that the two 
squeezed vacuum reservoirs are uncorrelated. The parameters 
$N_A, N_B, M_A$ and $ M_B$  describe the effects of squeezing of the reservoir 
modes. 
Actually, the parameters $N$ and $M$ 
represent the mean photon number and the phase property of the reservoirs,  
respectively, and are related as $|M|^{2} = N(N+1)$. 
Furthermore, introducing the density of states $g(\omega)$, where 
\begin{eqnarray*}
\sum_j \lambda_{j}\lambda_{2j_{a}-j} \rightarrow\int_0^{\infty}\!d\omega\, 
g(\omega)\lambda(\omega)\lambda(2\omega_{a}-\omega),
\end{eqnarray*} 
and  setting $t-t^\prime = \tau, $ one can easily show that
\begin{eqnarray}
\int_0^t\!dt^\prime\,e^{ \pm i(\omega_a -\omega)(t-t^\prime)}=\int_0^t\!d\tau\,
e^{\pm i(\omega_a-\omega)\tau }.\label{9} 
\end{eqnarray}
Since the exponential is a rapidly decaying function of time, the upper limit 
of integration can be extended to infinity. Making use of the approximate 
relation 
\begin{eqnarray}
 \int_0^{\infty}\!e^{\pm i(\omega_a -\omega)\tau} = \pi\,\delta(\omega_a-
\omega), \label{10} 
\end{eqnarray}
and applying the property of the Dirac delta function to the integrals of 
Eq.~(\ref{6}), we get 
\begin{eqnarray} \pi g(\omega_a) \lambda^{2}(\omega_a)=\frac{ {\gamma}_A }{2},
\label{11} 
\end{eqnarray} 
where ${\gamma}_A =2\pi g(\omega_a) \lambda^{2}(\omega_a)$ is the cavity 
damping constant for mode $A$. Similarly one can also show that the cavity 
damping constant for mode $B$ is given by 
${\gamma}_B =2\pi g(\omega_b) \lambda^{2}(\omega_b)$.   

In view of Eqs.~(7-11), after evaluating lengthy but straightforward 
consecutive integrations, Eq.~(\ref{6}) takes the form 
\begin{eqnarray}\frac{d\hat{\rho}(t)}{dt} &=& 
\frac{{\gamma}_A}{2}(N_A+1)\Big[2\,\hat{a} 
\hat{\rho}(t)\hat{a}^{\dag} -\hat{a}^{\dag}\hat{a}\hat{\rho}(t)- 
\hat{\rho}(t)\hat{a}^{\dag}\hat{a} \Big]  + \frac{{\gamma}_{A}N_{A}}
{2}\Big[2\,\hat{a}^{\dag}\hat{\rho}(t)\hat{a}- 
\hat{a}\hat{a}^{\dag}\hat{\rho}(t)-
\hat{\rho}(t)\hat{a}\hat{a}^{\dag}\Big] \nonumber\\
 & & +\frac{{\gamma}_{A}M_{A}}{2}\Big[2\,\hat{a}^{\dag}\hat{\rho}(t)
\hat{a}^{\dag} +2\,\hat{a}\hat{\rho}(t)\hat{a}-\hat{a}^{{\dag}2}\hat{\rho}(t)
 -\hat{\rho}(t)\hat{a}^{{\dag}2}- 
\hat{a}^{2}\hat{\rho}(t)-\hat{\rho}(t)\hat{a}^{2}\Big]\nonumber\\
 & & +\frac{{\gamma}_B}{2}(N_B+1)\Big[2\,\hat{b} 
\hat{\rho}(t)\hat{b}^{\dag} -\hat{b}^{\dag}\hat{b}\hat{\rho}(t) 
 -\hat{\rho}(t)\hat{b}^{\dag}\hat{b}\Big]
+\frac{{\gamma}_{B}N_{B}}{2}\Big[2\,\hat{b}^{\dag}\hat{\rho}(t)\hat{b}- 
\hat{b}\hat{b}^{\dag}\hat{\rho}(t)-
\hat{\rho}(t)\hat{b}\hat{b}^{\dag}\Big]\nonumber\\
 & &  +\frac{{\gamma}_{B}M_{B}}{2}\Big[2\,\hat{b}^{\dag}
\hat{\rho}(t)\hat{b}^{\dag} 
+2\,\hat{b}\hat{\rho}(t)\hat{b}-\hat{b}^{{\dag}2}\hat{\rho}(t)
 -\hat{\rho}(t)\hat{b}^{{\dag}2} 
  -\hat{b}^{2}\hat{\rho}(t)-\hat{\rho}(t)\hat{b}^{2}\Big].  \label{12}
\end{eqnarray}

In the cavity we consider two-modes of light known as the signal and idler 
modes produced by the  NDPO. The cavity has one single-port mirror in which 
light can enter or leave through while its other side is a mirror through 
which light may enter but can not leave. 
In this system we assume that a strong pump light of frequency $\omega_{0}$ 
interacts with a nonlinear-medium (crystal) inside the cavity and gives rise 
to a two-mode squeezed light (the signal-idler modes) with frequencies 
$\omega_a$ and $\omega_b$ such 
that $\omega_0=\omega_a+\omega_b$. With the pump mode treated classically 
(the amplitude of the pump mode is assumed to be real and constant), 
the interaction of the system is described, in the interaction picture, by the 
Hamiltonian 
\begin{eqnarray} 
\hat{H}_s = i\hbar\kappa{\gamma}_0 \big(\hat{a}\hat{b}-\hat{a}^{\dag}
\hat{b}^{\dag}\big), \label{13} 
\end{eqnarray}
where $ \kappa $ is the coupling constant and $\gamma_0 $ is the amplitude of 
the pump mode. Hence the master equation for the NDPO coupled to USVR, 
in view of Eqs.~(\ref{12}) and ~(\ref{13}), takes the form 
\begin{eqnarray}\frac{d\hat{\rho}(t)}{dt} & = & 
-\kappa{\gamma}_0 \Big[\hat{a}\hat{b}\hat{\rho}(t)-\hat{\rho}(t)\hat{a}\hat{b}+
\hat{\rho}(t)\hat{a}^{\dag}\hat{b}^{\dag} -
\hat{a}^{\dag}\hat{b}^{\dag}\hat{\rho}(t) \Big]\nonumber\\
 &  & + \frac{{\gamma}_A}{2}(N_A+1)\Big[2\,\hat{a} 
\hat{\rho}(t)\hat{a}^{\dag} -\hat{a}^{\dag}\hat{a}\hat{\rho}(t)- 
\hat{\rho}(t)\hat{a}^{\dag}\hat{a} \Big]  + \frac{{\gamma}_{A}N_{A}}
{2}\Big[2\,\hat{a}^{\dag}\hat{\rho}(t)\hat{a}- 
\hat{a}\hat{a}^{\dag}\hat{\rho}(t)-
\hat{\rho}(t)\hat{a}\hat{a}^{\dag}\Big] \nonumber\\
 &  & + \frac{{\gamma}_{A}M_{A}}{2}\Big[2\,\hat{a}^{\dag}\hat{\rho}(t)
\hat{a}^{\dag} +2\,\hat{a}\hat{\rho}(t)\hat{a}-\hat{a}^{{\dag}2}\hat{\rho}(t)
 -\hat{\rho}(t)\hat{a}^{{\dag}2}- 
\hat{a}^{2}\hat{\rho}(t)-\hat{\rho}(t)\hat{a}^{2}\Big]\nonumber\\
 &  & + \frac{{\gamma}_B}{2}(N_B+1)\Big[2\,\hat{b} 
\hat{\rho}(t)\hat{b}^{\dag} -\hat{b}^{\dag}\hat{b}\hat{\rho}(t) 
 -\hat{\rho}(t)\hat{b}^{\dag}\hat{b}\Big]
+\frac{{\gamma}_{B}N_{B}}{2}\Big[2\,\hat{b}^{\dag}\hat{\rho}(t)\hat{b}- 
\hat{b}\hat{b}^{\dag}\hat{\rho}(t)-
\hat{\rho}(t)\hat{b}\hat{b}^{\dag}\Big]\nonumber\\
 &  &  +\frac{{\gamma}_{B}M_{B}}{2}\Big[2\,\hat{b}^{\dag}
\hat{\rho}(t)\hat{b}^{\dag} 
+2\,\hat{b}\hat{\rho}(t)\hat{b}-\hat{b}^{{\dag}2}\hat{\rho}(t)
 -\hat{\rho}(t)\hat{b}^{{\dag}2} 
  -\hat{b}^{2}\hat{\rho}(t)-\hat{\rho}(t)\hat{b}^{2}\Big]  \label{14}
\end{eqnarray} 
This equation is the basis of our analysis and describes 
the interactions inside the cavity as well the interaction of the signal-idler 
modes produced by the NDPO and the squeezed vacuum 
reservoirs via the partially transmitting mirror. This master equation is 
consistent to that given in Ref.~[7] except that the expression there is for 
a single mode in a cavity coupled to a single mode vacuum reservoir.

\section{The Fokker-Planck Equation}

In this section we derive the Fokker-Planck equation for the Q-function. 
In order to obtain the 
Fokker-Planck equation for the Q-function corresponding to the master 
equation~(\ref{14}), one has first to put all terms in normal order.
Applying the commutation relations
\begin{subequations}
\begin{align}
 \Big[\hat{a},f(\hat{a},\hat{a}^{\dag})\Big] = &\frac{ \partial 
f(\hat{a},\hat{a}^{\dag})} { \partial\hat{a}^{\dag} }, \\
\Big[ \hat{a}^{\dag}, f(\hat{a},\hat{a}^{\dag}) \Big] = & -\frac{
\partial f(\hat{a},\hat{a}^{\dag}) } { \partial\hat{a} },       
\end{align}
\label{15}
\end{subequations}
one can verify that 
$\hat{a}\hat{\rho}=\hat{\rho}\hat{a}+\frac{ \partial\hat{\rho}}
{\partial\hat{a}^{\dag} },                                     $
$ \hat{\rho} \hat{a}^{\dag} =\hat{a}^{\dag} \hat{\rho}+\frac{
\partial\hat{\rho}}{\partial\hat{a} },                         $ 
where the density operator
$ \hat{\rho} =\hat{\rho} (\hat{a},\hat{a}^{\dag},t ) $
is considered to be in normal order. Making use of 
Eqs.~(\ref{15}), the relation 
$ [\hat{a},\hat{b}\hat{c}]=\hat{b}[\hat{a},\hat{c}]
+[\hat{a},\hat{b}]\hat{c} $ and the photonic commutation relation 
$ \hat{a}\hat{a}^{\dag} = \hat{a}^{\dag}\hat{a} + 1, $ the master equation 
(\ref{14}) can be written as 
\begin{eqnarray}\frac{d\hat{\rho}(t)}{dt} & = & -\kappa{\gamma}_0 \Big[
\frac{\partial\hat{\rho}}{\partial\hat{b}^{\dag}}\hat{a}+
\hat{a}^{\dag}\frac{\partial\hat{\rho}}{\partial\hat{b}}+
\hat{b}^{\dag}\frac{\partial\hat{\rho}}{\partial\hat{a}}+
\frac{\partial\hat{\rho}}{\partial\hat{a}^{\dag}}\hat{b}+
\frac{\partial^{2}\hat{\rho}}{\partial\hat{a}\partial\hat{b}}+ 
\frac{\partial^{2}\hat{\rho}}{\partial\hat{a}^{\dag}
\partial\hat{b}^{\dag}} \Big] \nonumber\\
&  & + \frac{{\gamma}_A}{2}(N_A+1) \Big[\frac{\partial }{ \partial\hat{a} }
(\hat{\rho}\hat{a}) +\frac{\partial}{ \partial \hat{a}^{\dag} }
( \hat{a}^{\dag} \hat{\rho} )+ 2 \frac{ \partial^{2} \hat{\rho} }
{ \partial \hat{a} \partial \hat{a}^{\dag} } \Big] - 
 \frac{{\gamma}_{A} N_{A}}{2} \Big[\frac{\partial}{ 
\partial\hat{a}^{\dag} }( \hat{a}^{\dag} \hat{\rho} )+
\frac{\partial}{ \partial \hat{a} }( \hat{\rho}\hat{a} ) \Big] \nonumber\\
&  &  -\frac{ {\gamma}_A }{2} M_A \Big[ \frac{\partial^{2}\hat{\rho}}
{\partial\hat{a}^{2} } +
\frac{\partial^{2}\hat{\rho} }{\partial\hat{a}^{{\dag}2} }\Big] \nonumber\\
&  & +  \frac{{\gamma}_B}{2}(N_B+1) \Big[\frac{\partial }{ 
\partial\hat{b} }(\hat{\rho}\hat{b}) +\frac{\partial}{ \partial 
\hat{b}^{\dag} }( \hat{b}^{\dag} \hat{\rho} )+ 2 \frac{ \partial^{2} 
\hat{\rho} }{ \partial \hat{b} \partial \hat{b}^{\dag} } \Big]
-\frac{{\gamma}_{B} N_{B}}{2}  \Big[\frac{\partial}{ 
\partial\hat{b}^{\dag} }( \hat{b}^{\dag} \hat{\rho} )+\frac{\partial}{ 
\partial \hat{b} }( \hat{\rho}\hat{b} ) \Big]\nonumber\\
&  & - \frac{ {\gamma}_B }{2} M_B \Big[ \frac{\partial^{2}\hat{\rho} }
{\partial\hat{b}^{2} }+
\frac{\partial^{2}\hat{\rho} }{\partial\hat{a}^{{\dag}2} } \Big]. \label{16}
\end{eqnarray}
In order to transform this equation into a c-number Fokker-Planck 
equation for the Q-function, one needs to multiply it on the left by  
$ \langle\alpha, \beta \mid $ and on the right by 
$ \mid \alpha, \beta\rangle,$ so that 
\begin{eqnarray} \frac{\partial Q}{\partial t} & = &
\Big[\kappa\gamma_{0} \big(\frac{\partial^{2}}{\partial
  \alpha \partial \beta}+
  \frac{\partial^{2}}{\partial \alpha ^{*}\partial\beta^{*}}+
  \frac{\partial}{\partial \beta^{*}}\alpha +
  \frac{\partial}{\partial \beta}\alpha^{*}  
  +\frac{\partial}{\partial\alpha^{*}}\beta+
  \frac{\partial}{\partial \alpha}\beta^{*}\big) \nonumber\\
&  & +  \gamma_{A}(N_A+1)\frac{\partial^{2}}
{\partial\alpha\partial\alpha^{*}}
 +\frac{\gamma_{A}}{2}\big(\frac{\partial }{\partial\alpha}\alpha+
\frac{\partial }{\partial \alpha^{*}}\alpha^{*}\big) -
\frac{\gamma_{A} M_A }{2}\big(\frac{\partial^{2}}{\partial
\alpha^{2}} +
\frac{\partial^{2}}{\partial\alpha^{{*}2} }\big)\nonumber\\
&  &  + 
\gamma_{B}(N_B+1)\frac{\partial^{2}}
{\partial\beta\partial\beta^{*}} + \frac{\gamma_{B}}{2}\big(\frac{\partial }{\partial \beta}\beta+
\frac{\partial }{\partial \beta^{*}}\beta^{*}\big)
  -  \frac{\gamma_{B} M_B}{2}\big(\frac{\partial^{2}}{\partial\beta^{2}} +
\frac{\partial^{2}}{\partial\beta^{{*}2}}\big)
 \Big] Q,                                          \label{17}
\end{eqnarray}
where 
\begin{eqnarray*}
Q=Q(\alpha^{*},\alpha,\beta^{*},\beta,t)=   
 \frac{1}{\pi^{2}}{\langle \alpha,\beta \mid \hat{\rho}(\hat{a}^{\dag},\hat{a},
\hat{b}^{\dag},\hat{b},t) \mid \alpha,\beta \rangle}. 
\end{eqnarray*} 
Expression~(\ref{17}) is the Fokker-Planck equation for the 
Q-function for the signal-idler modes produced by the NDPO coupled to two USVR.
To obtain the solution of this equation, we introduce the Cartesian 
coordinates defined by 
\begin{eqnarray}
 \alpha = x_1 + iy_1, 
\alpha^{*} =  x_1 - iy_1,  
\beta = x_2 + iy_2,  
\beta^{*} = x_2 - iy_2, \label{18} 
\end{eqnarray}
and note that 
\begin{eqnarray} x_1=\frac{1}{2}(\alpha + \alpha^{*}), 
y_1=-i\frac{1}{2}(\alpha - \alpha^{*}),
 x_2=\frac{1}{2}(\beta + \beta^{*}), y_2=-i\frac{1}{2}(\beta - \beta^{*}). 
\label{19}  
\end{eqnarray}
One can  show that 
\begin{subequations}
\begin{align} 
 \frac{\partial}{\partial \alpha} =  \frac{1}{2} 
\big( \frac{\partial}{\partial x_1 }
 - i\frac{\partial}{ \partial y_1 } \big),\\
\frac{\partial}{\partial\beta}  =  \frac{1}{2}\big(\frac{\partial}
{\partial x_2 } - i\frac{\partial}{\partial y_2 }\big). 
\end{align}
\label{20} 
\end{subequations}
Thus combining these results and their complex conjugates,  one 
readily obtains
\begin{eqnarray} 
\frac{\partial Q}{\partial t} &  =  & \Big[ \frac{\kappa\gamma_{0}}{2}
\Big(\frac{\partial^{2}}{\partial x_1 \partial x_2 } -
\frac{\partial^{2}}{\partial y_1 \partial y_2 }\Big) +\kappa\gamma_{0}
\Big(\frac{\partial}{\partial x_1}x_2 + \frac{\partial}{\partial x_2}x_1-
\frac{\partial}{\partial y_1}y_2- 
\frac{\partial}{\partial y_2} y_1\Big)\nonumber\\
&  &  + \frac{\gamma_{A}(N_A+1)}{4}\Big(\frac{\partial^{2}}{\partial x^{2}_1}
- \frac{\partial^{2}}{\partial y^{2}_1 }\Big) -\frac{\gamma_A M_A}{4}
\Big(\frac{\partial^{2}}{\partial x^{2}_1 } -
\frac{\partial^{2}}{\partial y^{2}_1 }\Big) \nonumber\\
&  & 
+\frac{\gamma_{B}(N_B+1)}{4}
\Big(\frac{\partial^{2}}{\partial x^{2}_2}
-\frac{\partial^{2}}{\partial y^{2}_2 }\Big)  -\frac{\gamma_B M_B}{4}
\Big(\frac{\partial^{2}}{\partial x^{2}_2 } -
\frac{\partial^{2}}{\partial y^{2}_2 }\Big) \Big]Q,      \label{21}
\end{eqnarray}
where $ Q = Q(x_1, x_2, y_1, y_2, t). $

Next, introducing the transformation defined by
$   x_1=x+u, \:\: x_2=x-u, \:\: y_1=y+v, \:\: y_2=v-y, $
one can  verify that
\begin{subequations}
\begin{align} 
& x=\frac{1}{2}(x_1+x_2),  s
& u= \frac{1}{2}(x_1-x_2),\\
& y=\frac{1}{2}(y_1-y_2),  
& v= \frac{1}{2}(y_1+y_2). 
\end{align}
\label{22}
\end{subequations} 
In view of these relations, it follows that
\begin{subequations} 
\begin{align}
& \frac{\partial}{\partial x_1}  =  \frac{1}{2}\Big[\frac{\partial}
{\partial x}+ \frac{\partial}{\partial u}\Big],
&\frac{\partial}{\partial x_2}=\frac{1}{2}\Big[\frac{\partial}
{\partial x}- \frac{\partial}{\partial u}\Big] \\
&\frac{\partial}{\partial y_1}  =  \frac{1}{2}\Big[\frac{\partial}
{\partial y}+ \frac{\partial}{\partial v}\Big],
&\frac{\partial}{\partial y_2}=\frac{1}{2}\Big[\frac{\partial}
{\partial v}- \frac{\partial}{\partial y}\Big]. 
\end{align}
\label{23}
\end{subequations}
Making use of Eqs.~(22, 23) in Eq.~(\ref{21}) 
and setting $\gamma_A = \gamma_B = \gamma $, $ N_A=N_B=N $ and $M_A=M_B=M$, 
for convenience, one arrives at
\begin{eqnarray}
\frac{\partial Q}{\partial t} & = & \Big[
\frac{\kappa\gamma_0+\gamma(N-M+1)}{8}\frac{\partial^{2}}{\partial x^{2}} +
\frac{\kappa\gamma_0+\gamma(N+M+1)}{8}\frac{\partial^{2}}
{\partial y^{2}} - \frac{\kappa\gamma_0-\gamma(N-M+1)}{8}\,\frac{\partial^{2}}
{\partial u^{2}} \nonumber\\
& & - \frac{\kappa\gamma_0-\gamma(N+M+1)}{8}\frac{\partial^{2}}
{\partial v^{2}} + \frac{2\kappa\gamma_0+\gamma}{2} \big(\frac{\partial}
{\partial x}x + \frac{\partial}{\partial y}y \big) 
- \frac{2\kappa\gamma_0-\gamma}{2}\big( \frac{\partial}{\partial u}u +
\frac{\partial}{\partial v} v \big) \Big]Q, \label{24}
\end{eqnarray}
which is the Fokker-Planck equation for the Q-function where $ Q=Q(x,y,u,v,t)$.

\section{ Solution of the Fokker-Planck Equation}

In this section  the explicit expression for the 
Q-function that describes the optical system is derived. In order to solve the 
differential equation (\ref{24}) using the propagator method 
discussed in Ref.~[1], one needs to transform the above equation into a 
Schr\"{o}dinger-type equation. This can be achieved upon replacing 
$ \big(\frac{\partial}{\partial x}, \frac{\partial}{\partial y},
\frac{\partial}{\partial u}, \frac{\partial}{\partial v}, x, y, u, v \big)$ 
and $ Q(x, y, u, v, t) $ by  $(i\hat{p}_x, i\hat{p}_y, i\hat{p}_u, 
i\hat{p}_v, \hat{x},\hat{y},\hat{u},\hat{v})$ and $ \mid Q(t)\rangle $ 
respectively. 
Hence Eq.~(\ref{24}) can be expressed as 
\begin{eqnarray} 
i\frac{d \mid Q(t)\rangle }{dt} &  = & i\Big[
-\frac{\lambda_1 }{8}\hat{p}^{2}_x
-\frac{\lambda_2 }{8}\hat{p}^{2}_y 
+\frac{\lambda_3 }{8}\hat{p}^{2}_u
+\frac{\lambda_4 }{8}\hat{p}^{2}_v 
+i\frac{\lambda_5 }{2}(\hat{p}_x \hat{x} +\hat{p}_y \hat{y} )
-i\frac{\lambda_6 }{2}(\hat{p}_u \hat{u} +\hat{p}_v \hat{v} )\Big]
\mid Q(t)\rangle \nonumber\\
&  = &  \hat{H}\mid Q(t)\rangle,      \label{25} 
\end{eqnarray}
where
\begin{subequations}
\begin{align} 
 &\lambda_{1,2}  =  \kappa\gamma_0+\gamma(N \mp M+1),   \\
 &\lambda_{3,4}  =  \kappa\gamma_0-\gamma(N \mp M+1),   \\
 &\lambda_{5,6}  =  2\kappa \gamma_0\pm\gamma.  
\end{align}
\label{26}
\end{subequations} 

A formal solution of Eq.~(\ref{25}) can be put in the form
\begin{eqnarray}
\mid Q(t)\rangle = \hat{u}(t)\mid Q(0)\rangle,
\label{27}
\end{eqnarray} 
where $\hat{u}(t)=exp(-i\hat{H}t/hbar)$ 
is a unitary operator and
\begin{eqnarray} 
\hat{H} =-i\frac{\lambda_1 }{8}\hat{p}^{2}_x
-i\frac{\lambda_2 }{8}\hat{p}^{2}_y +i\frac{\lambda_3 }{8}\hat{p}^{2}_u
+i\frac{\lambda_4 }{8}\hat{p}^{2}_v 
 -\frac{\lambda_5 }{2}(\hat{p}_x \hat{x} +\hat{p}_y 
\hat{y} ) +\frac{\lambda_6 }{2}(\hat{p}_u \hat{u} +
\hat{p}_v \hat{v})  \label{28}
\end{eqnarray}
is a quadratic quantum Hamiltonian.
Multiplying (\ref{27}) by $ \langle x,y,u,v \mid $ on the left yields
\begin{eqnarray} Q(x,y,u,v,t) = \langle x,y,u,v\mid\hat{u}(t)\mid Q(0)\rangle,
\label{29}  
\end{eqnarray}
where $ Q(x,y,u,v,t) = \langle x,y,u,v\mid Q(t)\rangle. $
Introducing a four-dimensional completeness relation for the position 
eigenstates 
$ \hat{I}=\int\!dx^{\prime}\,dy^{\prime}\,du^{\prime}\,dv^{\prime}\mid
x^{\prime}, y^{\prime}, u^{\prime}, v^{\prime}\rangle \langle
x^{\prime}, y^{\prime}, u^{\prime}, v^{\prime}\mid $
in expression (\ref{29}), one can see that
\begin{eqnarray}
 Q(x,y,u,v,t)=\int\!dx^{\prime}\,dy^{\prime}\,du^{\prime}\,dv^{\prime}\,
Q(x,y,u,v,t|  x^{\prime},y^{\prime},u^{\prime},v^{\prime},0) Q_{o}(x^{\prime},
y^{\prime},u^{\prime},v^{\prime}), \label{30} 
\end{eqnarray}
where
\begin{eqnarray}
 Q_{0}(x^{\prime},y^{\prime},u^{\prime},v^{\prime})= \langle x^{\prime},
y^{\prime},u^{\prime},v^{\prime}| Q(o)\rangle \label{31}  
\end{eqnarray}
is the initial Q-function and
$ Q(x,y,u,v,t|x^{\prime},y^{\prime},u^{\prime},v^{\prime},0) =\langle x,y,u,v|
\hat{u}(t)| x^{\prime},y^{\prime},u^{\prime},v^{\prime}\rangle  $
is the Q-function propagator.

Following Fesseha [1], the propagator associated with a quadratic 
Hamiltonian of the form
\begin{eqnarray}
 \hat H(\hat{x}_1,...,\hat{x}_n,\hat{p}_1,...,\hat{p}_n,t)=
\sum_{i=1}^{n}\Big[a_i \hat{p}^{2}_i+b_{i}(t)\hat{p}_i \hat{x}_i 
 +  c_{i}(t)\hat{x}_{i}^{2} \Big]    \label{32} 
\end{eqnarray}
is expressible as
 \begin{eqnarray} 
Q(x_{1},...,x_{n},t|x_{1}^{\prime},...,x_{n}^{\prime},0)=\Big[\frac{i}{2\pi}\Big]
 ^{\frac{n}{2}}\prod _{j=1}^{n}\sqrt{\frac{\partial ^{2}S_{c}}
  {\partial x_{j}\partial x_{j}^{\prime}}}\,\exp\Big[-\xi\int_{0}^{t}b_{j}
(t^{\prime})dt^{\prime}+ iS_{c}\Big], \label{33}  
\end{eqnarray}
where $S_{c}$ is the classical action, $\xi$ is a parameter related with 
operator ordering and $a_{i}$ is constant different from zero for the Hamiltonian to remain quadratic.
Comparing Eqs.~(\ref{32}) and~(\ref{28}), it follows that
$a_{i}= a_{x,y,u,v}=-\frac{i}{8}\lambda_{1,2,3,4}$,  
$(x_{1},x_{2},x_{3},x_{4})=
(x,y,u,v),$ $ c_{x}=c_{y}=c_{u}=c_{v}=0$, $b_{x}=b_{y}=-\frac{\lambda_5}{2}$, 
$b_{u}=b_{v}=\frac{\lambda_6}{2} $
and the  antistandard operator ordering $\xi =\frac{1}{2}.$
Thus the Q-function propagator associated with the Hamiltonian ~(\ref{28})
is expressible as
 \begin{eqnarray} 
Q(x,y,u,v,t|x^{\prime},y^{\prime},u^{\prime},v^{\prime},0)=
\frac{1}{4\pi^{2}}\Bigg[\frac{\partial ^{2}S_{c}}{\partial 
x^{\prime}\partial x}\frac{\partial ^{2}
S_{c}}{\partial y^{\prime}\partial y}
 \frac{\partial ^{2}S_{c}}
{\partial u{\prime}\partial u}\frac{\partial ^{2}S_{c}}
{\partial v^{\prime}\partial v}\Bigg]^{\frac{1}{2}}e^{iS_{c}+
  \frac{(\lambda_5-\lambda_6)}{2} t}. \label{34} 
\end{eqnarray} 
In order to obtain the explicit form of this expression, one has first 
to determine the classical action. To this end,  the Hamiltonian function 
corresponding to the quantum Hamiltonian (\ref{28}) is given by 
\begin{eqnarray} 
H =-i\frac{\lambda_1 }{8}p^{2}_x
-i\frac{\lambda_2 }{8}p^{2}_y +i\frac{\lambda_3 }{8}p^{2}_u
+i\frac{\lambda_4 }{8}p^{2}_v -
\frac{\lambda_5 }{2}(p_x x +p_y y ) +
\frac{\lambda_6 }{2}(p_u u +p_v v). \label{35}
\end{eqnarray} 
With the help of the Lagrangian
$ L= \sum_{i}\dot{x}_i p_i -H $
and the Hamilton equations
$ \dot{x}_i = \frac{\partial H}{\partial p_i}$ ($i=x, y, u, v$)
one can readily show that
\begin{eqnarray}  
L = \frac{2i}{\lambda_1}\big(\dot{x}+\frac{\lambda_5}{2}x\big)^{2}+
     \frac{2i}{\lambda_2}\big(\dot{y}+\frac{\lambda_5}{2}y\big)^{2} - 
\frac{2i}{\lambda_3}\big(\dot{u}-\frac{\lambda_6}{2}u\big)^{2}-
     \frac{2i}{\lambda_4}\big(\dot{v}-\frac{\lambda_6}{2}v\big)^{2}. \label{36}
\end{eqnarray} 
Applying the Euler-Lagrange equations
\begin{eqnarray}  
\frac{d}{dt}\Big(\frac{\partial L}{\partial \dot{x}_i}\Big) -
\frac{\partial L}{\partial x_i} = 0,  \label{37}
\end{eqnarray} 
along with Eq.~(\ref{36}), leads to
\begin{eqnarray*}
 \ddot{x}-\big(\frac{\lambda_5}{2}\big)^{2}x  =0,&   &   
 \ddot{y}-\big(\frac{\lambda_5}{2}\big)^{2}y =0,     \\
 \ddot{u}-\big(\frac{\lambda_6}{2}\big)^{2}u  =0,&   &  
 \ddot{v}-\big(\frac{\lambda_6}{2}\big)^{2}v =0. 
\end{eqnarray*}
The solutions of these differential equations can be written as
\begin{subequations}
\begin{align} 
& x(t) =a_1 e^{\frac{\lambda_5}{2}t}+ a_2 e^{-\frac{\lambda_5}{2}t},  &
& y(t) =b_1 e^{\frac{\lambda_5}{2}t}+b_2 e^{-\frac{\lambda_5}{2}t},  &   \\
&  u(t) =c_1 e^{\frac{\lambda_6}{2}t}+ c_2 e^{-\frac{\lambda_6}{2}t},  &
& v(t) =d_1 e^{\frac{\lambda_6}{2}t}+ 
d_2 e^{-\frac{\lambda_6}{2}t}. 
\end{align}
\label{38}  
\end{subequations}
Now substituting these expressions and their corresponding first order time 
derivatives
into Eq.~(\ref{36}), the Lagrangian takes the form
\begin{eqnarray*} L=2i\lambda^{2}_5\bigg(\frac{a^{2}_1}{\lambda_1} +
\frac{b^{2}_1}{\lambda_2}\bigg) e^{\lambda_5 t}-
2i\lambda^{2}_6 \bigg(\frac{c^{2}_2}{\lambda_3} +
\frac{d^{2}_2}{\lambda_4}\bigg) e^{-\lambda_6 t}. 
\end{eqnarray*}
On account of the above result, the classical action defined by
$ S_c=\int_{0}^{T}\!L(t)dt $
takes the form
\begin{eqnarray} 
S_c=2i\lambda^{2}_5\Big(\frac{a^{2}_1}{\lambda_1} +
\frac{b^{2}_1}{\lambda_2}\Big)\Big(e^{\lambda_5 T}-1\Big)+ 
2i\lambda^{2}_6\Big(\frac{c^{2}_2}{\lambda_3}+
\frac{d^{2}_2}{\lambda_4}\Big)\Big(e^{-\lambda_6 T}-1 \Big).\label{39}
\end{eqnarray} 
Applying the boundary conditions $x_{i}(0) =x^{\prime}_{i} $ and $x_{i}(T)=
x^{\prime\prime}_{i}$ in Eq.~(\ref{38}), one can obtain that
\begin{eqnarray*}
 a_1 = \frac{x^{\prime\prime}e^{\frac{\lambda_5}{2}T}-x^{\prime}}
  {e^{\lambda_5 T}-1},&    &    
 b_1 = \frac{y^{\prime\prime}e^{\frac{\lambda_5}{2}T}-y^{\prime}}
  {e^{\lambda_5 T}-1},  \\ 
 c_2  = \frac{u^{\prime\prime} e^{\frac{-\lambda_6}{2}T}-u^{\prime}}
  {e^{-\lambda_6 T}-1},&   &   
 d_2 = \frac{v^{\prime\prime}e^{\frac{-\lambda_6}{2}T}-v^{\prime}}
  {e^{-\lambda_6 T}-1}. 
\end{eqnarray*}  
Inserting the above expressions into Eq.~(\ref{39}) and replacing 
$ (x^{\prime\prime},
y^{\prime\prime}, u^{\prime\prime},
v^{\prime\prime}, T) $ by  $ (x,y,u,v,t) $ yields 
\begin{eqnarray} 
S_c=2i\lambda_5 \bigg[
\frac{\big(x^{\prime}- e^{ \frac{\lambda_5}{2}t} \big)^{2} }
{\lambda_1\big(e^{\lambda_5 t}-1\big)}+
\frac{\big(y^{\prime}- e^{\frac{\lambda_5}{2}t} \big)^{2} }
{\lambda_2 \big(e^{\lambda_5 t}-1 \big)}\bigg] 
+ 2i\lambda_6 \bigg[\frac{\big(u^{\prime}- e^{\frac{-\lambda_6}
{2}t}\big)^{2} }{\lambda_3
\big(e^{-\lambda_6 t}-1\big)}+
\frac{\big(v^{\prime}- e^{\frac{-\lambda_6}{2}t}\big)^{2} }
{\lambda_4 \big(e^{-\lambda_6 t}-1\big)} \bigg] \label{40}
\end{eqnarray} 
and  employing this relation the following results are obtained:
\begin{subequations}
\begin{align}  
& \frac{ \partial^{2} S_c }{ \partial x \partial x^{\prime} }=
-\frac{4i\lambda_{5}\,e^{ \frac{ \lambda_5}{2}t} }
{\lambda_1 \big( e^{\lambda_5 t}-1 \big) }, & 
&  \frac{ \partial^{2} S_c }{ \partial y \partial y^{\prime} }=
-\frac{4i\lambda_{5}\,e^{ \frac{ \lambda_5}{2}t} }
{\lambda_2 \big( e^{\lambda_5 t}-1 \big) }, &\\  
& \frac{ \partial^{2} S_c }{ \partial u \partial u^{\prime} }=
-\frac{4i\lambda_{6}\,e^{-\frac{\lambda_6}{2}t}}
{\lambda_3 \big( e^{-\lambda_6 t}-1 \big) }, & 
& \frac{ \partial^{2} S_c }{ \partial v \partial v^{\prime} }=
-\frac{4i\lambda_{6} e^{-\frac{\lambda_6}{2}t}}
{\lambda_4 \big( e^{-\lambda_6 t}-1 \big) }. 
\end{align}
\label{41}
\end{subequations}
Thus, in view of Eq.~(\ref{41}), the Q-function propagator (\ref{34}) takes the form
\begin{eqnarray} 
 Q(x,y,u,v,t|x^{\prime},y^{\prime},u^{\prime},v^{\prime},0) & = &
 \frac{4\lambda_{5}\lambda_{6}}{\pi^{2}\sqrt{\lambda_{1}\lambda_{2}\lambda_{3}
\lambda_{4}}} \frac{\,e^{(\lambda_{5}-
\lambda_{6})t}}{ (e^{\lambda_{5}t}-1)(e^{-\lambda_{6} t}-1)} \nonumber\\
& & \times\exp\bigg[-\frac{2\lambda_{5}}
{(e^{\lambda_{5}t}-1)}
 \Big(\frac{ x^{\prime 2}-2x\,e^{\frac{\lambda_{5}}{2}t} x^{\prime}
+x^{2}e^{\lambda_5\,t}}{\lambda_1}\nonumber\\
& &+\frac{ y^{\prime 2}-2y\,e^{\frac{\lambda_{5}}{2}t} y^{\prime} + 
y^{2}e^{\lambda_{5}t}}{\lambda_{2}}\Big) \nonumber\\
& & -\frac{2\lambda_{6}}{(e^{-\lambda_{6}t}-1)}
 \Big(\frac{ u^{\prime 2}-2u \,e^{\frac{-\lambda_{6}}{2}t} u^{\prime}
+u^{2}e^{-\lambda_{6}t}}{\lambda_{3}} \nonumber\\
& & + \frac{ v^{\prime 2}-2v\,e^{\frac{-\lambda_{6}}{2}t} v^{\prime} + 
v^{2}e^{-\lambda_{6}t}}
{\lambda_{4}}\Big) \bigg]. \label{42}
\end{eqnarray}  
 
Considering the signal-idler modes produced by the NDPO 
to be initially in a two-mode vacuum state, the initial Q-function 
 is expressible as
\begin{eqnarray*}
 Q_{0}(\alpha^{\prime},\beta^{\prime})=
\frac{1}{\pi^{2}}\langle \alpha^{\prime},\beta^{\prime}| 0,0 \rangle \langle 0,0|\alpha^{\prime},\beta^{\prime}\rangle = 
\exp(-\alpha^{\prime *}\alpha^{\prime}- \beta^{\prime *}\beta^{\prime}), 
 \end{eqnarray*}
and in terms of the Cartesian variables of expression~(\ref{19}), 
this equation becomes
\begin{eqnarray*}
Q_{0}(x^{\prime}_1,x^{\prime}_2,y^{\prime}_1,y^{\prime}_2) =\frac{1}{\pi^{2}}
\exp\Big[-\big( x^{\prime 2}_1+x^{\prime 2}_2+
y^{\prime 2}_1+y^{\prime 2}_2\big)\Big]. 
\end{eqnarray*}
Furthermore, in terms of $x^{\prime},y^{\prime},u^{\prime}$ and $ v^{\prime},$
one can write
\begin{eqnarray*}
 \int\!dx^{\prime}_{1}\,dx^{\prime}_{2}\,dy^{\prime}_{1}\,dy^{\prime}_{2}\,
Q_{0}(x^{\prime}_{1},x^{\prime}_{2}, y^{\prime}_{1},y^{\prime}_{2})=
\int\!dx^{\prime}\,dy^{\prime}\,du^{\prime}\,dv^{\prime}\,
Q_{0}(x^{\prime},y^{\prime}, u^{\prime},v^{\prime}),
\end{eqnarray*}
where
\begin{eqnarray*}
 Q_{0}(x^{\prime},y^{\prime}, u^{\prime},v^{\prime})=\frac{|J|}{\pi^2}
\exp\Big[-2\big( x^{{\prime} 2}+y^{{\prime} 2}+u^{{\prime} 2}+v^{{\prime} 2}
\big)\Big] 
\end{eqnarray*}
and $ J $ is the Jacobian of the transformation of $x_1$, $x_2$, $y_1$ and $y_2$
with respect to $x$, $y$, $y$ and $v$.
Making use of Eq.~(\ref{19}) in the Jacobian, one can  show that $|J|=4$. 
Hence
\begin{eqnarray} 
Q_{0}(x^{\prime},y^{\prime}, u^{\prime},v^{\prime})=\frac{4}{\pi^2}\,\exp\Big[
-2\big( x^{{\prime}2}+y^{{\prime}2}+u^{{\prime}2}+v^{{\prime}2}
\big)\Big]. \label{43}
\end{eqnarray}
Substituting expression~(\ref{41}) into Eq.~(\ref{34}) and then combining 
the result with Eq.~(\ref{43}) and finally carrying out the integration in Eq.~(\ref{30}) 
applying the relation
\begin{eqnarray*}
\int_{-\infty}^{\infty}\!dx^{\prime}\,\exp\Big[-k x^{{\prime}2}+
d x^{\prime}\Big]=
\sqrt{\frac{\pi}{k}}\,\exp\Big[ \frac{d^{2}}{4k} \Big],\:\:\:\:\:k>0,   
\end{eqnarray*}
the Q-function takes the compact form
\begin{eqnarray}
 Q(x,y,u,v,t) = \frac{4}{\pi^{2} \sqrt{ a_{1} a_{2} a_{3} a_{4}}}
\exp\Big[-\frac{2}{ a_1}x^{2} -\frac{2}{a_2}y^{2} -\frac{2}{a_3}u^{2}
-\frac{2}{a_4}v^{2} \Big], \label{44}
\end{eqnarray}
where
\begin{subequations}
\begin{align}
& a_{1,2}=\frac{\lambda_{1,2} \big(e^{\lambda_{5}t}-1\big)+\lambda_{5}}
{\lambda_{5} e^{\lambda_{5}t}}, \\
&  a_{3,4}=\frac{\lambda_{3,4} \big(e^{-\lambda_{6}t}-1\big)+\lambda_{6}}
{\lambda_{6} e^{-\lambda_{6}t}}. 
\end{align}
\label{45} 
\end{subequations}
It can be easily verified that the Jacobian of the inverse transformation is
$|J^{\prime}| =\frac{1}{4}$. One can then write
\begin{eqnarray*}
\int\!dx\,dy\,du\,dv\,Q(x,y,u,v,t)=\int\!dx_{1}\,dx_{2}\,dy_{1}\,dy_{2}\,
Q^{\prime}(x_1,x_2,y_1,y_2,t), 
\end{eqnarray*}
in which the final expression for $ Q^{\prime}(x_1,x_2,y_1,y_2,t) $
is obtained from Eq.~(\ref{44}) employing the inverse 
transformations ~(\ref{22}).
Upon carrying out further inverse transformations ~(\ref{19}),  the required 
final form the Q-function for the signal-idler modes produced by the 
nondegenerate parametric oscillator ~(NDPO) coupled to two uncorrelated 
squeezed vacuum reservoirs takes the form
\begin{eqnarray} 
Q(\alpha,\alpha^{*},\beta,\beta^{*},t) & = & \frac{D}{\pi^{2}}\exp\Big[-b_{1}
(|\alpha|^2 +|\beta|^2 ) + 
b_{2}(\alpha\beta+\alpha^{*}\beta^{*}) +
b_{3}(\alpha\beta^{*}+\alpha^{*}\beta) \nonumber\\
& & +\frac{b_4}{2}(\alpha^{2}+\alpha^{*2}+
\beta^{2}+\beta^{*2})\Big], \label{46}
\end{eqnarray}
where
\begin{subequations}
\begin{align}
& D  =  \frac{1}{\sqrt{ a_{1} a_{2}a_{3} a_{4}}},\\
& b_{1,2}  = \frac{1}{4}\bigg[\frac{1}{\pm a_{1}}\pm \frac{1}{a_{2}}+ 
\frac{1}{a_{3}}+\frac{1}{a_{4}}\bigg],\\
& b_{3,4}  =  \frac{1}{4}\bigg[-\frac{1}{a_{1}}+\frac{1}{a_{2}}\pm 
\frac{1}{a_{3}}\mp
\frac{1}{a_{4}}\bigg].
\end{align}
\label{47}
\end{subequations}
This Q-function is useful to calculate the expectation values of antinormally 
ordered operators and consequently the quadrature variances. It could also be 
used to calculate the photon number distribution of different optical systems.
In this paper, this function is used to calculate the quadrature fluctuations 
(variances) of the NDPO coupled to two USVR. It can be readily verified that 
the Q-function (\ref{46}) is positive and normalised.

Now we proceed to obtain the expressions for the Q-function for some
special cases of interest: For the case when there are no squeezed vacuum 
reservoirs ($r=0$), that is, when the external 
environment is an ordinary vacuum, the Q-function (\ref{46}) takes the form 
\begin{eqnarray}
 Q(\alpha,\alpha^{*},\beta,\beta^{*},t)=\frac{1}{\pi^{2}a_{1}a_{3}}
\exp\bigg[-\frac{1}{2}\Big(\frac{a_1+a_3}{a_{1}a_{3}}\Big)
(|\alpha|^2 + |\beta|^2)+\frac{1}{2}\big(\frac{a_1-a_3}
{a_{1} a_{3}}\big)(\alpha\beta+\alpha^{*}\beta^{*})\bigg]. \label{48} 
\end{eqnarray}
This is the Q-function for the nondegenerate parametric oscillator coupled to 
ordinary vacuum. On the other hand, in the absence of damping ($\gamma=0$), 
Eq.~(\ref{46}) reduces to the form
\begin{eqnarray}
 Q(\alpha,\alpha^{*},\beta,\beta^{*},t)=\frac{{\rm sech}\kappa\gamma_{0}t}
{\pi^{2}} \exp\Big[-|\alpha|^2-|\beta|^2- 
(\tanh\,\kappa\gamma_{0}t)
(\alpha\beta+\alpha^{*}\beta^{*})\Big], \label{49}
\end{eqnarray}
which  is the Q-function for the nondegenerate parametric amplifier.

Next we obtain the Q-function for the single-mode generated by a degenerate 
parametric oscillator coupled to a single-mode squeezed vacuum reservoir from
the Q-function for the NDPO (\ref{46}). 
The Q-function for the single-mode can be expressed as 
\begin{eqnarray*}
Q(\alpha,\alpha^{*},t)=
\int\!d^{2}\beta\,Q(\alpha,\alpha^{*},\beta,\beta^{*},t),
\end{eqnarray*}
so that using Eq.~(\ref{46}) and the relation
\begin{eqnarray}
 \int\!\!\!d^{2}\alpha\exp\big[\!-a^{\prime}|\alpha|^2+\!b^{\prime}\alpha\!\! +\!\!
c^{\prime}\alpha^{*}\!\! +\! A^{\prime}\alpha^{2}\!+
\!B^{\prime}\alpha^{*2}\big]\!
=\!\frac{1}{\sqrt{\big({a^{\prime}}^{2\!\!}-\!\! 4A^{\prime}B^{\prime}\big)}}
\!\!\!\exp\Big[\frac{a^{\prime}b^{\prime}c^{\prime}+A^{\prime} 
{c^{\prime}}^{2}\!+\!B^{\prime} b^{2}}
{{a^{\prime}}^{2} \!- \!4A^{\prime}B^{\prime}}\Big], a^{\prime}>0 
\label{50}
\end{eqnarray}
the Q-function for the DPO coupled to a single-mode squeezed vacuum reservoir
takes the form 
\begin{eqnarray}
 Q(\alpha,\alpha^{*},t)=\frac{D}{\pi\sqrt{y}}
\exp\bigg[-a |\alpha|^2+
\frac{A}{2}\Big(\alpha^{2}+\alpha^{*2}\Big)\bigg],\label{51}
\end{eqnarray}
where
\begin{subequations}
\begin{align}
& y  =  b^{2}_1 -b^{2}_4,  \\
& a  =  \frac{1}{y}\Big[ (b_1+b_4)\big(b_{1}(b_1-b_4)+2b_{2}b_{3}\big) -
b_{1}(b_2+b_3)^{2} \Big], \\
& A  = \frac{1}{y}\Big[(b_1+ b_{4})\big(b_4(b_1-b_4)+2b_{2}b_{3}\big)+b_{4}
\big(b_2+ b_3\big)^{2} \Big].
\end{align}
\label{52}
\end{subequations}
Upon integrating Eqs.~(\ref{48}) and~(\ref{49}) with respect to $\beta$ by 
employing relation~(\ref{50}), one can also find the Q-function for the 
DPO in the absence of  squeezed vacuum reservoir ($r=0$)  and in the 
absence of damping ($\gamma = 0$)  to be
\begin{eqnarray*} 
Q(\alpha,\alpha^{*},t)=\frac{2}{\pi (a_{1}+a_{3})}
\exp\Big[-\frac{2}{a_1+a_3}|\alpha|^2\Big] 
\end{eqnarray*}
and 
\begin{eqnarray}
 Q(\alpha,\alpha^{*},t)=\frac{ {\rm sech}^{2}\,\kappa\gamma_{0}t }{\pi} 
 \exp\big[-({\rm sech}^{2}\,\kappa\gamma_{0}t)(|\alpha|^2)\big], 
\label{53}
\end{eqnarray}
respectively.

\section{ Quadrature Squeezing}

In this section the intracavity quadrature fluctuations for 
the single-mode generated by the DPO as well as the signal-idler modes 
produced by the NDPO coupled to the two squeezed vacuum reservoirs using the 
pertinent Q-functions derived in the previous section are analysed. 

Here the first focus is the squeezing properties of the single-mode light. 
These properties could be described by two Hermitian operators defined as 
$ \hat{a}_1=\hat{a}^{\dag}+\hat{a}$ and  
$\hat{a}_2=i(\hat{a}^{\dag}-\hat{a}). $
These quadrature operators obey the commutation relation 
$[\hat{a}_1,\hat{a}_2]=2i.  $
The variance of these quadrature operators can be put in the form 
\begin{eqnarray}
\big(\Delta\hat{a}_{1,2}\big)^{2}&=&\langle{\hat{a}_{1,2}}^2\rangle-
{\langle\hat{a}_{1,2}\rangle}^2 
\label{54}  
\end{eqnarray}

We now proceed to calculate the expectation values involved in expression 
(\ref{54}). Applying the relation 
\begin{eqnarray}
\langle\hat{A}(\hat{a},\hat{a}^{\dag})\rangle=
\int_{-\infty}^{\infty}\!d^{2}\alpha\,Q(\alpha,\alpha^{*},t) 
A_{a}(\alpha,\alpha^{*}),\label{55}  
\end{eqnarray}
in which $A_{a}(\alpha,\alpha^{*})$ is the c-number equivalent of the operator 
$\hat{A}(\hat{a},\hat{a}^{\dag}) $ for the antinormal ordering, one arrives at
\begin{eqnarray*}
\langle\hat{a}\rangle=\int_{-\infty}^{\infty}\!d^{2}\alpha\,Q(\alpha,
\alpha^{*},t)\,\alpha. 
\end{eqnarray*}
Upon using the Q-function (\ref{51}) for the single mode, the 
above equation can be expressed as
\begin{eqnarray*} 
\langle\hat{a}\rangle=\frac{D}{\sqrt{y}}\left.\frac{\partial}
{\partial b}\int_{-\infty}^{\infty}\frac{d^{2}\alpha}{\pi}
 \exp\Big[-a\alpha^{*}
\alpha+ \frac{A}{2}\big(\alpha^{2}+\alpha^{*2}\big)+b\alpha 
\Big]\right |_{b=0},
\end{eqnarray*}
and on the basis of ~(\ref{50}) for which $c^{\prime}=0$ and $A^{\prime}=
B^{\prime}$, 
one can verify that
\begin{eqnarray*}
 \langle\hat{a}\rangle=\frac{D}{\sqrt{y}}\left.\frac{\partial}{\partial b}
\Bigg[\frac{\exp\Big(\frac{A b^{2} }{2(a^{2}-A^{2})}\Big)}
{\sqrt{a^{2}-A^{2}}}\Bigg]\right |_{b=0} =0.  
\end{eqnarray*}
In view of this result expression (\ref{55}) reduces to
\begin{eqnarray}
\big(\Delta\hat{a}_{1,2}\big)^{2}=1+2\langle\hat{a}^{\dag}\hat{a}\rangle \pm
\langle\hat{a}^{{\dag}2}\rangle \pm \langle\hat{a}^{2}\rangle.\label{56}
\end{eqnarray}
Making use of the fact that the c-number equivalent of $\hat{a}^{\dag}\hat{a}$
for the antinormal ordering is $ \alpha^{*}\alpha-1 $ and 
applying relation ~(\ref{55}) in evaluating all the expectation  values in 
Eq.~(\ref{56}), we arrive at
\begin{eqnarray}
\big(\Delta\hat{a}_{1,2}\big)^{2}=\frac{2}{a \mp A}-1. \label{57}
\end{eqnarray}

Finally the quadrature fluctuations of the single-mode at any time $t$, in 
view of Eqs.~(\ref{52}),~(\ref{46}) and~(\ref{45}), take the form
\begin{eqnarray*} 
\big(\Delta\hat{a}_{1,2}\big)^{2}=\frac{\lambda_{1,2} \big(e^{\lambda_{5}t}
-1\big)+\lambda_{5}}{\lambda_{5} e^{\lambda_{5}t}}+
\frac{\lambda_{3,4} \big(e^{-\lambda_{6}t}-1\big)+\lambda_{6}}
{\lambda_{6} e^{-\lambda_{6}t}}-1.  
\end{eqnarray*}
At steady-state $({t \rightarrow \infty}),$ the variances given above 
reduce to
\begin{eqnarray*}
\big(\Delta\hat{a}_{1,2}\big)^{2}=\frac{\lambda_{1,2}}{\lambda_{5}}+
\frac{\lambda_{3,4}}{\lambda_{6}}-1,
\end{eqnarray*}
and with the aid of Eq.~(\ref{26}) one can rewrite these expressions as
\begin{eqnarray}
\big(\Delta\hat{a}_{1,2}\big)^{2}=
\frac{2(N \mp M)+1}{1-\big(\frac{2\kappa\gamma_{0}}
{\gamma}\big)^{2}}.\label{58}
\end{eqnarray}
Since for  squeezed vacuum reservoirs
\begin{subequations}
\begin{align} 
& N  = \sinh^{2}r, \\
& M  = \sinh r \cosh r , 
\end{align}
\label{59}
\end{subequations}
where $r$ is the squeezing parameter taken to be real and positive 
for convenience, expression~(\ref{59}) takes the form 
\begin{eqnarray} 
\big(\Delta\hat{a}_{1,2}\big)^{2}=\frac{e^{\mp 2r}}
{1-\big(\frac{2\kappa\gamma_{0}}{\gamma}\big)^{2}}.\label{60}
\end{eqnarray}
Using (\ref{60}) one can show that
$(\Delta\hat{a}_1)^{2}<1$, for 
\begin{eqnarray} 
r > -\frac{1}{2}ln\bigg[1-\Big(\frac{2\kappa\gamma_{0}}
{\gamma}\Big)^{2}\bigg] 
\label{61}
\end{eqnarray}
and 
$(\Delta\hat{a}_2)^{2} > 1$
for all $r$.
This shows that the degenerate parametric oscillator coupled to a squeezed 
vacuum reservoir is in a squeezed state  for the value of $ r $ specified by 
Eq.~(\ref{61}).

In the absence of squeezing, i.e., $r=0$, substitution of Eq.~(\ref{26}) 
into Eq.~(\ref{57}) leads to
\begin{eqnarray*} 
(\Delta\hat{a}_1)^{2}=(\Delta\hat{a}_2)^{2} = \frac{1-\frac{\kappa
\gamma_{0}}{\gamma} e^{-(\gamma-2\kappa\gamma_{0})t}}{\Big[1-
\big(\frac{2\kappa\gamma_{0}}{\gamma}\big)^{2}\Big]} 
\Bigg[\big(1 - e^{-4\kappa\gamma_{0}t}\big)+
\frac{2\kappa\gamma_{0}}{\gamma}\big(1+e^{-4\kappa\gamma_{0}t}\big)\Bigg].
\end{eqnarray*}
At steady-state and when the parametric oscillator is operating below 
threshold 
($ \gamma > 2\kappa\gamma_{0}$), 
this equation reduces to
\begin{eqnarray*} 
(\Delta\hat{a}_1)^{2}=(\Delta\hat{a}_2)^{2}=\frac{1}{\Big[1-
(\frac{2\kappa\gamma_{0}}{\gamma})^{2}\Big]}
\end{eqnarray*} 
in which both variances become greater than unity. Hence the single-mode in 
this case is not in a squeezed state.

In the absence of damping ($\gamma = 0$), Eq.~(\ref{57}) reduces to
\begin{eqnarray*} 
(\Delta\hat{a}_1)^{2}=(\Delta\hat{a}_2)^{2}=2\overline{n} + 1,
\end{eqnarray*}
where
$ \overline{n}=\sinh^{2}\kappa\gamma_{0}t $ is the mean photon number 
for the single-mode. From these variances one can infer that 
the single-mode in this case is in a chaotic state as expected.
Furthermore, in the absence of parametric interaction ($\kappa=0 $), 
Eq.~(\ref{57}) could be expressed as
\begin{eqnarray*} 
(\Delta\hat{a}_{1,2})^{2}=2a_{1,2} -1 =1-\Big[1- e^{-\gamma t}\Big]
\Big[1\mp e^{\mp 2r}\Big]\lessgtr 1, 
\end{eqnarray*}
and at steady-state these relations reduce to
\begin{eqnarray}
(\Delta\hat{a}_{1,2})^{2} = e^{\mp 2r},\label{62}
\end{eqnarray}
which are the quadrature fluctuations of the squeezed vacuum reservoir $A$.

Now we proceed to investigate the 
squeezing properties of the signal-idler modes produced by the NDPO 
coupled to the two squeezed vacuum reservoirs applying the 
Q-function ~(\ref{46}). 
The squeezing properties of  two-mode light can be described by two 
quadrature operators defined as 
\begin{eqnarray}
 \hat{c}_{1,2} =\frac{1}{\sqrt{2}}\Big(\hat{a}_{1,2} +\hat{b}_{1,2} \Big),
\label{63}
\end{eqnarray}
where
\begin{subequations}
\begin{align}
 \hat{a}_1 =(\hat{a}^{\dag} +\hat{a}), &  & 
 \hat{b}_1 =(\hat{b}^{\dag} +\hat{b}), \\  
 \hat{a}_2 =i(\hat{a}^{\dag} -\hat{a}), &  &  
\hat{b}_2 =i(\hat{b}^{\dag} -\hat{b}), 
\end{align}
\label{64}
\end{subequations}
and $\hat{a}\,(\hat{b})$ denotes the annihilation operator for the intracavity 
mode $a\,(b)$. The quadrature operators $\hat{c}_1$ and $\hat{c}_2$ satisfy 
the commutation relation $[\hat{c}_1,\hat{c}_2]=2i$.
On account of these expressions, the variances can be expressed as
\begin{eqnarray}
\big(\Delta\hat{c}_{1,2}\big)^{2} &=&
\langle\hat{c}^{2}_{1,2}\rangle-{\langle\hat{c}_{1,2}\rangle}^{2}\nonumber \\
&=&\frac{1}{2}\langle\hat{a}^{2}_{1,2}\rangle+
\frac{1}{2}\langle\hat{b}^{2}_{1,2}\rangle +\langle \hat{a}_{1,2},
\hat{b}_{1,2} \rangle ,\label{65}
\end{eqnarray}
in which 
\begin{eqnarray*}
\langle \hat{a}_i,\hat{b}_i \rangle= \langle \hat{a}_i\hat{b}_i \rangle-
\langle \hat{a}_i\rangle\langle\hat{b}_i \rangle,
\end{eqnarray*}
and $i=1,2$.
In particular, when $a$ and $b$ represent the signal and idler modes, 
respectively,
it can be shown that
\begin{eqnarray}
 (\Delta\hat{c}_{1,2})^{2}&= &\frac{1}{2}(\Delta\hat{a}_{1,2})^{2}+
\frac{1}{2}(\Delta\hat{b}_{1,2})^{2} + \langle \hat{a}_{1,2},\hat{b}_{1,2}
\rangle \nonumber \\ 
&= & (\Delta\hat{a}_{1,2})^{2}+\langle \hat{a}_{1,2},\hat{b}_{1,2}\rangle
\label{66}
\end{eqnarray}
as $ (\Delta\hat{a}_{1,2})^{2}=(\Delta\hat{b}_{1,2})^{2}$ and  $\langle 
\hat{a}_{1,2}\rangle = \langle\hat{b}_{1,2}\rangle=\langle\hat{c}_{1,2}
\rangle=0 $.
In order to obtain the explicit form of Eq.~(\ref{66}), we proceed as 
follows. In view of expression ~(\ref{64}) and ~(\ref{55}), one can express 
that
\begin{eqnarray*}
\langle \hat{a}_1 \hat{b}_1 \rangle=
\int_{-\infty}^{\infty}\!d^{2}\alpha\, d^{2}\beta\,(\alpha^{*}+\alpha) 
(\beta^{*}+\beta)\, Q(\alpha,\alpha^{*},\beta^{*},
\beta,t).
\end{eqnarray*}
Then employing the Q-function~(\ref{46}) the above equation can be further 
expressed as
\begin{eqnarray*}
\langle \hat{a}_1 \hat{b}_1 \rangle &=&D \int_{-\infty}^{\infty}
\!\!\frac{d^{2}\alpha}{\pi}
(\alpha^{*}+\alpha)\,exp\big[-b_1 \alpha^{*}\alpha+
\frac{b_4}{2}(\alpha^{2}+\alpha^{*2}\big] \\
& & \times\,\int_{-\infty}^{\infty}\!\!\frac{d^{2}\beta}{\pi}(\beta^{*}+\beta)
\,exp\big[-b_{1}\beta^{*}\beta +(b_{2}\alpha+b_{3}\alpha^{*})\beta \\
& & +(b_{2}\alpha^{*}+b_{3}\alpha)\beta^{*}+
\frac{1}{2}b_{4}(\beta^{2}+\beta^{*2})\big].
\end{eqnarray*}
On setting $K=b_{2}\alpha+b_{3}\alpha^{*}$,
\begin{eqnarray*}
\langle \hat{a}_1 \hat{b}_1 \rangle &=& D 
\int_{-\infty}^{\infty}\frac{d^{2}\alpha}{\pi}
(\alpha^{*}+\alpha)\,exp\Big[-b_1 \alpha^{*}\alpha+
\frac{b_4}{2}(\alpha^{2}+\alpha^{*2})\Big]
\Big(\frac{\partial}{\partial K}+\frac{\partial}{\partial K^{*}}\Big)\\
& & \times\int_{-\infty}^{\infty}\frac{d^{2}\beta}{\pi} exp\Big[-b_{1}
\beta^{*}\beta +
(b_{2}\alpha+b_{3}\alpha^{*})\beta+(b_{2}\alpha^{*}+b_{3}\alpha)\beta^{*}+
\frac{1}{2}b_{4}(\beta^{2}+\beta^{*2})\Big],
\end{eqnarray*} 
so that performing the integration with respect to $\beta$ on the basis of 
relation~(\ref{50}) and carrying out the differentiation we obtain
\begin{eqnarray*}
\langle \hat{a}_1 \hat{b}_1 \rangle=\frac{D}{y^{\frac{3}{2}}}(b_1+b_4)(b_2+b_3)
\int_{-\infty}^{\infty}\frac{d^{2}\alpha}{\pi}(\alpha^{2}+
\alpha^{*2}+2\alpha^{*}\alpha)
exp\Big[-a\alpha^{*}\alpha+
\frac{A}{2}(\alpha^{2}+\alpha^{*2})\Big],
\end{eqnarray*}
from which it follows that
\begin{eqnarray*}
\langle \hat{a}_1\hat{b}_1 \rangle=\frac{D}{y^{\frac{3}{2}}}(b_1+b_4)(b_2+b_3)
\Big(2\frac{\partial}{\partial A}-
2\frac{\partial}{\partial a}\Big)\int_{-\infty}^{\infty}
\frac{d^{2}\alpha}{\pi}exp\Big[-a\alpha^{*}\alpha+
\frac{A}{2}(\alpha^{2}+\alpha^{*2})\Big].
\end{eqnarray*}
Next, integrating over $\alpha$ and carrying out the differentiation, we get
\begin{eqnarray*}
\langle \hat{a}_1 \hat{b}_1 \rangle=\frac{D}{y^{\frac{3}{2}}(b_1+b_4)
(b_2+b_3)}\Big(\frac{2A+2a}{a^{2}-A^{2}}\Big)\frac{1}{\sqrt{a^{2}-A^{2}}}.
\end{eqnarray*}
Making use of expression~(\ref{47}) along with Eq.~(\ref{52}) the above 
equation reduces to
\begin{eqnarray*}
\langle \hat{a}_1\hat{b}_1 \rangle=a_1 - a_3.
\end{eqnarray*}
A similar approach leads to 
\begin{eqnarray*}
\langle \hat{a}_2,\hat{b}_2 \rangle=-(a_2 - a_4).
\end{eqnarray*}
Now Eq.~(\ref{66}) can be put as
\begin{eqnarray*}
(\Delta\hat{c}_{1,2})^{2}=2a_{1,4} - 1,
\end{eqnarray*}
and at this stage the variances are given by
\begin{eqnarray*}
\big (\Delta\hat{c}_{1,2}\big)^{2}=2\Big[\frac{\kappa\gamma_{0} 
\pm \gamma(N \mp M+1)}{2\kappa\gamma_{0} \pm \gamma} \big(1-e^{ 
\mp (2\kappa\gamma_{0} \pm \gamma)t}\big) +
e^{ \mp (2\kappa\gamma_{0} \pm \gamma)t}\Big] - 1.
\end{eqnarray*}
Finally the quadrature fluctuations  of the signal-idler modes at any 
time $t$,  in view of Eq.~(\ref{59}), take the form
\begin{subequations} 
\begin{align}
&\big(\Delta\hat{c}_{1}\big)^{2}=1-\Big[1-e^{-(\gamma + 2\kappa\gamma_{0})t}
\Big]\Big[1- \frac{\gamma\,e^{-2r}}{\gamma + 2\kappa\gamma_{0}}\Big] < 1, \\
&\big(\Delta\hat{c}_{2}\big)^{2}=1-\Big[1-e^{-(\gamma - 2\kappa\gamma_{0})t}
\Big]\Big[1- \frac{\gamma\,e^{+2r}}{\gamma - 2\kappa\gamma_{0}}\Big]> 1.
\end{align}
\label{67}
\end{subequations}
Hence the signal-idler modes generated by the NDPO coupled
to the USVR, when operating below threshold 
($\gamma-2\kappa\gamma_{0} > 0$), are  in  squeezed states for all values of 
$r$. 

At steady-state (${t \rightarrow \infty}$), Eq.~(\ref{67}) can 
be put in the form
\begin{subequations} 
\begin{align}
&\big(\Delta\hat{c}_{1}\big)^{2}=\Big(\frac{\gamma}{\gamma 
+ 2\kappa\gamma_{0}}\Big) e^{- 2r} < 1, \\
&\big(\Delta\hat{c}_{2}\big)^{2}=\Big(\frac{\gamma}{\gamma 
- 2\kappa\gamma_{0}}\Big) e^{+ 2r} > 1.   
\end{align}
\label{68}
\end{subequations}
This equation clearly shows the possibility of a very large amount of squeezing
(approaching 100\%) below the standard quantum limit in the one quadrature 
at the expense of enhanced fluctuations in the other quadrature, where in this 
case the standard quantum limit is taken to be $\sqrt{(\Delta\hat{c}_1)^{2}}
\sqrt{ (\Delta\hat{c}_2)^{2}}=1$.
In addition, at threshold ($\gamma=2\kappa\gamma_{0}$), one  obtains
\begin{subequations} 
\begin{align}
&(\Delta\hat{c}_1)^{2} =\frac{1}{2}e^{-2r},\\
& (\Delta\hat{c}_2)^{2}\;  \rightarrow \infty .
\end{align}
\label{69}
\end{subequations}

In the absence of squeezed vacuum reservoirs $(r=0)$, expression (68) becomes 
\begin{eqnarray} 
 (\Delta\hat{c}_{1,2})^{2}=\frac{\gamma \pm \Big(2\kappa\gamma_{0}\,
e^{-(\gamma \pm 2\kappa\gamma_{0})t}\Big)}
{\gamma \pm 2\kappa\gamma_{0}} \lessgtr 1.\label{70} 
\end{eqnarray}
This shows that the signal-idler modes produced by the nondegenerate 
parametric oscillator in the absence of squeezed vacuum reservoirs are also in 
 squeezed states. At steady-state and at threshold, these relations reduce to 
\begin{eqnarray}
(\Delta\hat{c}_1)^{2} & = &  \frac{1}{2},\nonumber\\
(\Delta\hat{c}_2)^{2} \;& \rightarrow & \infty .\label{71}
\end{eqnarray}
In this case one can easily see that there is only a 50\% reduction of noise 
below the vacuum level.  By comparing Eqs.~(\ref{69}) 
and~(\ref{71}) we can conclude that coupling of the NDPO to the squeezed 
vacuum reservoirs is essential for the generation of a larger amount of 
squeezing.

In the absence of damping ($\gamma=0$), expression (\ref{67}) reduces to 
\begin{eqnarray}
\Delta\hat{c}^{2}_{1,2}= e^{\mp 2\kappa\gamma_{0}t} \lessgtr 1,\label{72}
\end{eqnarray} 
which are the quadrature fluctuations of the signal-idler modes produced 
by the nondegenerate parametric amplifier. This indicates that the 
nondegenerate parametric amplifier coupled to ordinary vacuum reservoirs 
also generates squeezed states. 

Finally, when there is no parametric interaction inside the cavity 
$(\kappa=0)$, Eq.~(\ref{67}) takes the form 
\begin{eqnarray}
\big(\Delta\hat{c}_{1,2}\big)^{2}=1-\big(1- e^{-\gamma t}
\big)\Big[1\mp e^{\mp 2r}\Big] \lessgtr 1,\label{73}
\end{eqnarray}
which, at steady-state, leads to 
\begin{eqnarray}
(\Delta\hat{c}_{1,2})^{2}= e^{\pm 2r},\label{74}
\end{eqnarray}
which are the quadrature fluctuations of the reservoir modes $A$ and $B$. 
Upon comparing the relations~(\ref{71}) and~(\ref{74}) with~(\ref{69}), one 
can see that the 
quadrature variances at steady-state and at threshold are the product of the 
variances of the NDPO coupled to ordinary vacuum and the variances 
pertaining to the squeezed vacuum reservoirs. Furthermore upon comparing 
expressions ~(\ref{62}) and ~(\ref{74}) one can observe that at steady-state 
the variances of a signal mode squeezed vacuum reservoir as well as those of 
two independent squeezed vacuum reservoirs are the same.

\section{ Conclusion} 
 
We have derived the master equation for the 
signal-idler modes produced by the nondegenerate parametric oscillator coupled 
to two uncorrelated squeezed vacuum reservoirs and consequently the 
Fokker-Planck equation. We have solved the pertinent 
Fokker-Planck equation which is a second order differential equation applying 
the propagator method ~\cite{28} and obtained a compact form of the Q-function 
of the optical system coupled to two independent squeezed vacuum reservoirs. 
We have also deduced the Q-functions for a NDPO coupled to ordinary vacuum 
reservoirs,  degenerate parametric oscillators coupled to a squeezed vacuum 
reservoir and an ordinary vacuum reservoir, and for the 
nondegenerate and degenerate parametric amplifiers from the Q-function for the 
NDPO coupled to the two USVR. 

In general the Q-function can be used to evaluate the expectation values of
antinormally ordered operators as well as photon number distributions 
for the NDPO and other similar optical systems. 

We have calculated the nonlinear quantum quadrature fluctuations of the 
signal-idler modes generated by a nondegenerate parametric oscillator below 
threshold coupled to two uncorrelated squeezed vacuum reservoirs, using the 
Q-function. Although it is a well known fact that quantum noise can not be 
eliminated, we have  shown that the signal-idler 
modes produced by the optical system are in a two-mode squeezed state at 
any time $t$. 
More interestingly, we have shown that at steady-state and below threshold
it is possible to generate an optimal squeezing in one of the quadratures 
below the standard quantum limit 
at the expense of enhanced fluctuations in the other quadrature so that 
the Heisenberg uncertainty principle remains valid.
Furthermore calculation of the quadrature fluctuations at threshold clearly 
shows that it is possible to produce an arbitrarily large squeezing 
(approaching 100\%) in one of the quadratures with an infinitely large
noise in the other quadrature. We have also shown that the degenerate 
parametric oscillator could be in a squeezed state for  a squeezing parameter 
above a certain value when it is coupled to a squeezed vacuum reservoir. 

We have shown that the coupling of the optical system to the squeezed vacuum 
reservoirs is essential in order to get a more suppressed noise in one of the 
quadratures. 

Finally we have calculated the quadrature fluctuations for the nondegenerate 
parametric amplifier coupled to ordinary vacuum reservoirs and verified that 
it also generates squeezed states.

\section*{Acknowledgements}

I would like to thank  K. Fesseha,  M. Lewenstein and A. Sanpera for fruitful 
discussions.  
I acknowledge financial support by the Deutscher Akademiker 
Austausch Dienst (DAAD).

\end{document}